\documentclass[a4paper,12pt]{article}

\usepackage{amsmath}
\usepackage{booktabs}
\usepackage{cite}
\usepackage{dcolumn}
\usepackage{graphicx}
\usepackage[latin1]{inputenc}
\usepackage{lscape}
\usepackage{psfrag}
\usepackage{setspace}
\usepackage{topcapt}
\usepackage{xspace}
\begin{document}

\title{Supercooled Lennard-Jones Liquids and Glasses: 
  a Kinetic Monte Carlo Approach}
\author{Javier Hern\'andez-Rojas \\
{\small Departamento de F\'\i sica Fundamental II, Universidad de La Laguna,} \\
{\small 38205 Tenerife, Spain} \\
        David J. Wales \\
  {\small University Chemical Laboratories,  Lensfield Road,} \\
  {\small Cambridge, CB2 1EW, UK}}


\maketitle

\hrule

\begin{abstract}
A kinetic Monte Carlo (KMC) method is used to study the structural 
properties and dynamics of a supercooled binary Lennard-Jones 
liquid around the glass transition temperature.  
This technique permits us to explore the potential energy surface 
and barrier distributions without 
suffering the exponential slowing down at low temperature that affects
molecular dynamics simulations.
In agreement with previous studies we observe a distinct change in behaviour around $T=0.45$, 
close to the dynamical transition temperature $T_c$ of mode coupling theory (MCT). 
Below this temperature the number of different local minima visited by the
system for the same number of KMC steps decreases by more than an order of magnitude.
The mean number of atoms involved in each jump 
between local minima and the average distance they move also decreases significantly, 
and new features appear in the partial structure factor.
Above $T\sim0.45$ the probability distribution for the magnitude of
the atomic displacement per KMC step exhibits an exponential decay, which is
only weakly temperature dependent.
\end{abstract}

\hrule

\newpage


\section{Introduction}
A complete microscopic theory to explain the complex phenomenology of
supercooled liquids and glasses is still elusive, and is the subject 
of a considerable research effort\cite{condmatt99}.
Mode-coupling theory, based upon a continuum approach to the wave vector dependence
of correlation functions, has proved quite successful for supercooled liquids
at higher temperatures, before activated dynamics must be accounted for explicitly\cite{gotze91,gotzeS92,kob97,gotze99}.
For finite systems, such as clusters, it has recently been possible to establish detailed
connections between structure,
thermodynamics, dynamics and the underlying potential energy surface (PES)\cite{WalesDMMW00}.
In the context of glasses, it was Goldstein who first noted the crucial role of the PES\cite{goldstein69},
and Stillinger and Weber who first analysed computationally the local minima (`inherent structures') of various
bulk models\cite{stillingerw82}. By analysing transitions between inherent
structures they were able to identify the slow barrier
crossings and localised rearrangements at low temperatures anticipated by Goldstein\cite{stillingerw83c}.

Computer simulation has played a significant role in the testing and development of models for
supercooled liquids and glasses\cite{Kob99}. Interest in more direct connections to the PES
has recently increased, with Sastry, Debenedetti and Stillinger\cite{sastryds98} characterising
`landscape-influenced' and `landscape-dominated' regimes for a binary Lennard-Jones system, in
agreement with the instantaneous normal modes picture of Donati, Sciortino and Tartaglia\cite{donatist00}.
A number of other studies have also begun to characterise features of the underlying PES in more
detail\cite{heuer97,AngelaniPRV98,buchnerh99,HeuerB00,AngelaniPRV00,AngelaniDRSS00,BroderixBCZG00,MiddletonW01}.
However, it has not yet been possible to account for the properties of supercooled liquids
and glasses at the level of detail achieved for finite systems\cite{WalesDMMW00}.



One obvious problem with the simulation of low temperature supercooled liquids
by standard molecular dynamics (MD) techniques is that the exploration of the PES
slows down exponentially with decreasing temperature.
In the present contribution we show how a kinetic Monte Carlo (KMC) approach
\cite{Voter86,FichthornW91} can provide useful information on the dynamics of a binary Lennard-Jones system around
the glass transition temperature. This technique combines accurate calculations of
elementary transition states and the associated pathways with approximate dynamics,
and hence it can overcome low temperature trapping and provide access to much longer
simulation time scales. Hence it provides a view complementary to mode-coupling theory and
standard MD simulations, since the rate theory employed is most accurate at low temperatures,
and loses its validity at high temperatures.
The paper is organised as follows. In section II we introduce the model and
discuss the relevant points of the KMC method used. In 
sections III and IV we present the details of the KMC approach that
we have used, and the numerical results,   
and in the last section we finish with the conclusions. 

\section{The Model}

The system we have studied is a binary mixture of $N=60$ atoms
in a cubic box with periodic boundary conditions, 
where $80\%$ of the atoms are of type $A$ and $20\%$ are of type $B$.
They interact via a Lennard-Jones potential of the form
\begin{equation}
V_{\alpha \beta}=4\epsilon_{\alpha \beta}  
\left[ \left( \frac{\sigma_{\alpha \beta}}{r_{\alpha \beta}} \right)^{12}-
\left( \frac{\sigma_{\alpha \beta}}{r_{\alpha \beta}} \right)^{6} \right],
\end{equation}
where $r_{\alpha \beta}$ is the distance between particles and 
$\alpha$,$\beta$ $\in$ $\left\{ A,B \right\} $.
The parameters we used were
$\epsilon_{AB}=1.5$, $\epsilon_{BB}=0.5$, 
$\sigma_{AB}=0.8$ and $\sigma_{BB}=0.88$, 
and all quantities will be measured in reduced units: energy in 
$\epsilon_{AA}$, length in $\sigma_{AA}$, temperature in 
$\epsilon_{AA}/k_B$  ($k_B$ is the Boltzmann constant)
and time in $(\sigma_{AA}^2 m/\epsilon_{AA})^{1/2}$, 
with $m$ the mass of both types of atom.
The density was fixed at $1.2$.
We employed the Stoddard-Ford shifted, truncated version of the above
potential\cite{Stoddardf73}, with a cutoff at $r_c=1.842$,
along with the minimum image convention \cite{All87}.

Binary systems such as this have proved very popular amongst the glass simulation community,
since they do not crystallise on the molecular dynamics (MD) time
scale\cite{koba94,koba95a,koba95b,sastryds98,buchnerh99,donatigpkp99,angellbv99,sciortinokt99,%
SchroderSDG00,donatist00,Sastry00a,Sastry00b,%
SciortinoKT00,AngelaniDRSS00,BroderixBCZG00,buchnerH00,sastry01,DebenedettiS01}.
Stillinger and Weber were probably the first to consider an 80:20 mixture in their
simulation of ${\mathrm{Ni_{80}P_{20}}}$ \cite{stillingerw85d}. 
The present parameterisation was introduced by Kob and Andersen \cite{koba94}
when they modified the potential of Ernst et al.~\cite{ernstng91} because they found
that it crystallised at low temperatures. The free energy barrier separating supercooled
liquid from crystal is sufficiently high for the present parameters that crystallisation
has not been reported in conventional MD studies\cite{MiddletonHMW01}.

Following Angell \cite{Angell95a,Angell97,GreenIXA99}, 
glass-forming liquids can be classified as strong or fragile. 
Fragile systems exhibit non-Arrhenius temperature dependence of
transport properties such as the diffusion constant, usually accompanied by
a significant heat capacity peak at the glass transition. In contrast, strong
systems exhibit Arrhenius dynamics and small or negligible changes in 
thermodynamic properties at the glass transition.
Previous work has shown that the present system exhibits a significant degree of 
non-Arrhenius behaviour at low temperatures \cite{koba94,koba95a,koba95b}.

Although the supercell size of only 60 atoms is 
rather small it has been shown in previous work to
retain many of the most important features of the supercooled state
\cite{BroderixBCZG00,BhattacharyaBKZ99,buchnerh99,AngelaniPRV00}.
Our aim in the present work is to analyse KMC trajectories consisting of a fixed number
of elementary rearrangements between local minima as a function of temperature. In this
preliminary study we report some basic structural and thermodynamic results, for comparison
with previous work, and focus in particular on the changes that occur in the rearrangements
sampled at lower temperatures. The energies of stationary points will be reported per atom,
but energy barriers will be reported per supercell, since they scale intensively, not extensively, with the
supercell size.

\section{The Kinetic Monte Carlo Approach} 

Within the landscape picture we are concerned with the set of all 
particle positions, ${\bf r}_i$ with  $i=1,2,\ldots,N$, which we can view as a 
$3N$-dimensional vector ${\bf R}$, and its hopping dynamics on the PES, $V({\bf R})$.
For our purposes the most important points on the PES are the
minima and the transition states that connect them.
Here we follow Murrell and Laidler and define a true transition state as a saddle point
with Hessian index one \cite{MurrellL68}, $i.e.$ the Hessian has only one negative eigenvalue
whose eigenvector corresponds to the reaction coordinate and connects two 
minima. True transition states are expected to dominate the dynamics in the low temperature regime 
because of the Murrell-Laidler theorem: if two minima are connected by a higher index saddle
then there must be a lower energy path mediated by one or more true transition states\cite{MurrellL68}.

The hopping dynamics between minima were simulated using a KMC approach \cite{Voter86,FichthornW91}. 
We started from an initial random configuration
and searched for a local minimum using 
Nocedal's limited memory LBFGS routine \cite{lbfgs}.
From the current local minimum we searched for 
transition states that connect it to other local minima. 
These searches were performed using
a hybrid BFGS/eigenvector-following algorithm \cite{MunroW99}
where diagonalisation of the Hessian matrix is avoided. In fact, the Hessian
is not required at all, but for the present empirical potential it proved
more efficient to calculate the uphill search direction from the Hessian than
to use a variational approach\cite{MunroW99}.
To find more than one transition state for each minimum we used
the method introduced by Middleton and Wales, where starting points for
transition state searches are proposed by considering hard-sphere-type moves
\cite{MiddletonW01}.
We employed a maximum of twenty transition states searches from each minimum. 
The minima connected to each transition state were then obtained from approximate
steepest-descent paths commencing parallel and antiparallel to the
transition vector using LBFGS energy minimisation\cite{lbfgs}. All stationary points 
were finally converged to a root-mean-square force of less than $10^{-6}$
using one or two eigenvector-following steps, employing full diagonalisation
of the analytic Hessian matrix, to ensure that the stationary points have
the correct Hessian index.

The rate constants, used to calculate the probability per unit time of jumping
to a new minimum, were calculated using Rice-Ramsperger-Kassel-Marcus (RRKM)
theory within the harmonic approximation \cite{GilbertS90,KunzB95}:
\begin{equation}
K_{j \leftarrow i}=\frac{g_i}{\bar{g}_{ij}}
\frac{\prod\limits_{\alpha=1}^{3N-3}\nu_{\alpha}^i}
     {\prod\limits_{\alpha=1}^{3N-4}\bar\nu_{\alpha}^{ij}} 
e^{-\Delta E/k_B T},
\end{equation}
\noindent
where $g_i$ and $\bar{g}_{ij}$ are the order of the symmetry group of minimum
$i$ and transition state $ij$ (which are both almost always unity for the present system);
$\nu_{\alpha}^i$ and
$\bar\nu^{ij}_\alpha$ are the frequencies of minimum $i$ and 
transition state $ij$; $\Delta E$ is the potential energy difference between minimum $i$
and the transition state, and $T$ is the temperature.

If all the rate constants out of minimum $i$ are known then we can evaluate the jump probability between two 
minima $i$ and $j$ as
\begin{equation}
\label{P}
P_{j \leftarrow i}=\frac{K_{j \leftarrow i}}
{\sum\limits_{k}K_{k \leftarrow i}},
\end{equation}
\noindent
where the sum is over all transition states. 
In the KMC method we randomly select one of the connected minima according to
the probabilities defined in Eq.~(\ref{P}). 
Finally, once the system has moved to a new minimum the process is repeated.

As in previous practical implementations of the KMC approach \cite{SnurrBT94},
a further approximation is implicit because the sampling of transition states
connected to minimum $i$ is likely to be incomplete. However, this error does
not affect the relative probabilities among the pathways located, but only the
estimate of the corresponding time scale, which is approximate in any case
because of the model rate constants employed. 
Furthermore, we expect the omitted transition states to correspond to
higher barrier processes with smaller rate constants on the basis of statistics
obtained from more extensive surveys of stationary points \cite{MiddletonW01}.
The application of KMC also carries
with it the assumption that the jumps between minima are uncorrelated, so that
the system equilibrates for long enough in each minimum to lose memory of the
previous trajectory. We expect this assumption to be well satisfied in the
low temperature regime below about $T=0.5$, which is our main concern in the present work.
As we noted above, the low temperature regime in which the KMC approach is most accurate
should be complementary to approaches such as mode-coupling theory and standard
simulation methods.

The KMC approach differs in several ways from the 
activation-relaxation technique (ART) of Barkema and Mousseau
et al.~\cite{BarkemaM96,BarkemaM98,BarkemaM99,BarkemaM00,BarkemaMD00,MousseauB00}. 
We calculate the transition states and pathways to very high precision, and locate
more than one transition state per minimum in order to implement the KMC procedure.
In previous ART studies the objective has generally been to achieve efficient structural
relaxation, not to simulate true dynamics, and a Metropolis criterion has been used
for jumps between minima using only the potential energy difference, without regard
for effects such as vibrational entropy.

\section{Results}
\begin{figure}[htbp]
\psfrag{-4.71}[Br][Br]{$-4.71$}
\psfrag{-4.66}[Br][Br]{$-4.66$}
\psfrag{-4.61}[Br][Br]{$-4.61$}
\psfrag{-4.56}[Br][Br]{$-4.56$}
\psfrag{-4.51}[Br][Br]{$-4.51$}
\psfrag{-4.46}[Br][Br]{$-4.46$}
\psfrag{-4.41}[Br][Br]{$-4.41$}
\psfrag{-4.36}[Br][Br]{$-4.36$}
\psfrag{-4.31}[Br][Br]{$-4.31$}
\psfrag{-4.26}[Br][Br]{$-4.26$}
\psfrag{-4.21}[Br][Br]{$-4.21$}
\psfrag{0}[tc][tc]{$0$}
\psfrag{1000}[tc][tc]{$1000$}
\psfrag{2000}[tc][tc]{$2000$}
\psfrag{3000}[tc][tc]{$3000$}
\psfrag{lowest minimum}[Bl][Bl]{lowest minimum}
\psfrag{T=1}[cl][cl]{$T=1.000$}
\psfrag{T=0.6}[cl][cl]{$T=0.600$}
\psfrag{T=0.5}[cl][cl]{$T=0.500$}
\psfrag{T=0.45}[cl][cl]{$T=0.450$}
\psfrag{T=0.425}[cl][cl]{$T=0.425$}
\psfrag{V/N}[Bc][tc]{$\left<V\right>_n$}
\psfrag{n}[tc][tc]{$n$}
\centerline{\includegraphics[width=14.0cm]{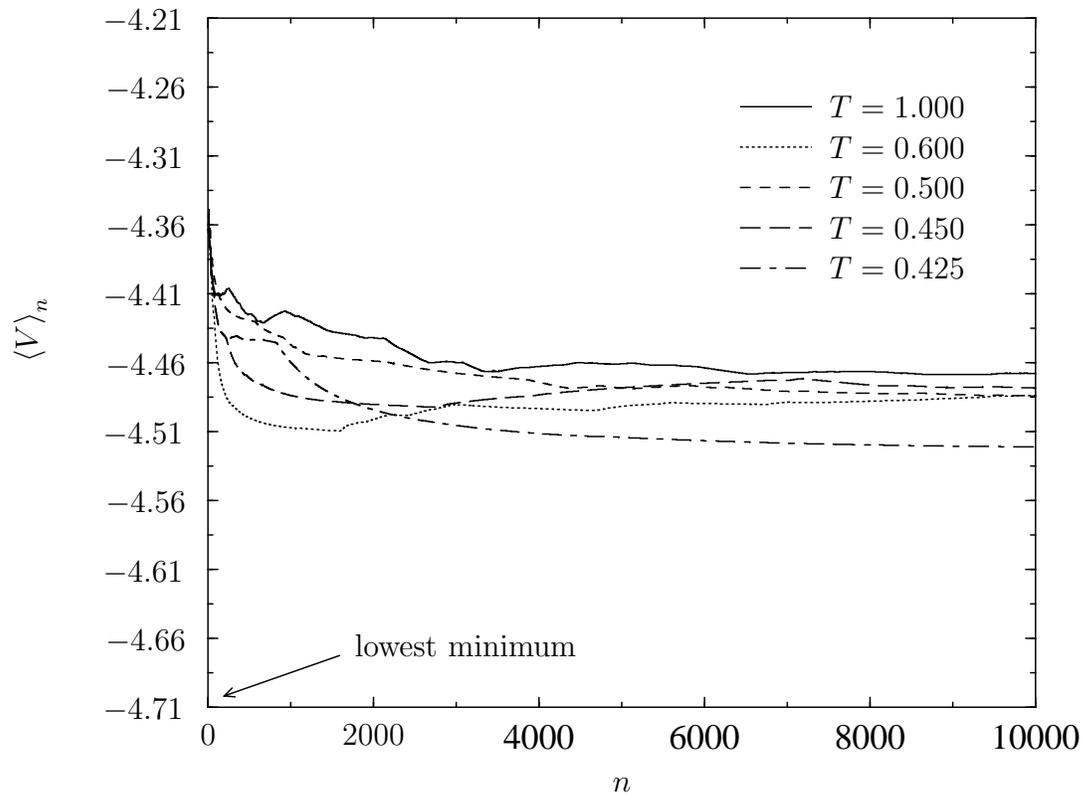}}
\caption{\label{fig:term} Running average of the potential energy per atom, $\left<V\right>_n$,
as a function of the number of KMC steps, $n$, at temperatures 
$T=1$, 0.6, 0.5, 0.45 and 0.425. 
We indicate with an arrow the energy of the lowest minimum that we have found.}
\end{figure}

We have performed a series of numerical KMC simulations to compute different properties
of the liquid as we vary the temperature. 
We used a random starting configuration in each case, and hence it is important to
discard the initial part of the trajectory when the system is still equilibrating.
In Figure \ref{fig:term} we plot
the running average energy per atom of the local minima at KMC step $n$, defined as
\begin{equation}
\left< V \right>_n = {1\over n}\sum_{j=1}^n V_j,
\end{equation}
where $V_j$ is the energy per atom of the local minimum at KMC step $j$.
In the course of a simulation, and before averaging other quantities, we have to be sure that 
at least local equilibration has been achieved. We assume that the system has
equilibrated sufficiently when 
$\left< V \right>_n$ reaches a stable value, which occurs after
about 5000 KMC steps for all temperatures (Figure \ref{fig:term}).
However, for the lowest temperature runs it is obvious that complete equilibration
has not been achieved, since the crystal should be the lowest free energy minimum\cite{MiddletonHMW01}.
On the other hand, at low temperature the system samples fewer minima 
for KMC runs consisting of equal numbers of steps,
and consequently the {\it local\/} equilibration actually appears to be faster. 

Our mean energies, averaged over the last 5000 KMC steps, 
are similar to those obtained in reference \cite{BroderixBCZG00} using
MD simulation (and a slightly different cut-off), except for our value at $T=1$, which appears to
be significantly too low. Similar behaviour has been observed before in comparison of ART and
MD results for silica, where the local minima obtained by ART sampling at the highest temperature
were rather lower than those obtained from MD. For the KMC case the failure at high temperature is
likely due to the breakdown in the Markov assumption for the minimum to minimum model
dynamics. The system does not stay long enough in each minimum to establish equilibrium properties
and lose its memory of previous jumps. The transition rate between high energy minima is therefore
probably underestimated, and a systematic bias towards lower energy minima results.

\begin{figure}[htbp]
\centerline{
\includegraphics[width=14.0 cm]{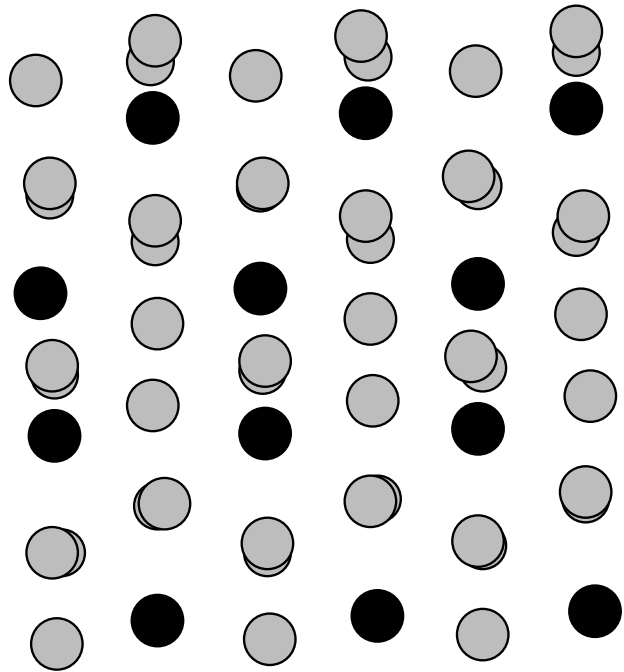}}
\caption{\label{fig:min_ene} Lowest minimum found for the
binary Lennard-Jones mixture. The 48 A atoms and 12 B atoms are shaded
gray and black, respectively. 
}
\end{figure}

The regions of the PES sampled in the present work are all high in potential energy
compared to the lowest minimum that we found for this system, which is shown in Figure \ref{fig:min_ene}.
In this crystalline structure, based on a distorted face-centred-cubic lattice,
the B atoms are arranged in a zig-zag pattern \cite{MiddletonHMW01}. This minimum 
(energy $-4.71$ per atom)
was found using a new stochastic global optimisation method, 
which combines the basin-hopping approach \cite{lis87,WalesD97a}
with KMC-type steps between local minima; further details will be published elsewhere.

\begin{figure}[htbp]
\psfrag{Different local minima sampled}[Bc][tc]{different local minima sampled}
\psfrag{0}[Br][Br]{$0$}
\psfrag{250}[Br][Br]{$250$}
\psfrag{500}[Br][Br]{$500$}
\psfrag{750}[Br][Br]{$750$}
\psfrag{1000}[Br][Br]{$1000$}
\psfrag{Temperature}[tc][tc]{temperature}
\psfrag{0.4}[tc][tc]{$0.4$}
\psfrag{0.5}[tc][tc]{$0.5$}
\psfrag{0.6}[tc][tc]{$0.6$}
\psfrag{0.7}[tc][tc]{$0.7$}
\psfrag{0.8}[tc][tc]{$0.8$}
\psfrag{0.9}[tc][tc]{$0.9$}
\psfrag{1}[tc][tc]{$1.0$}
\centerline{\includegraphics[width=14.0cm]{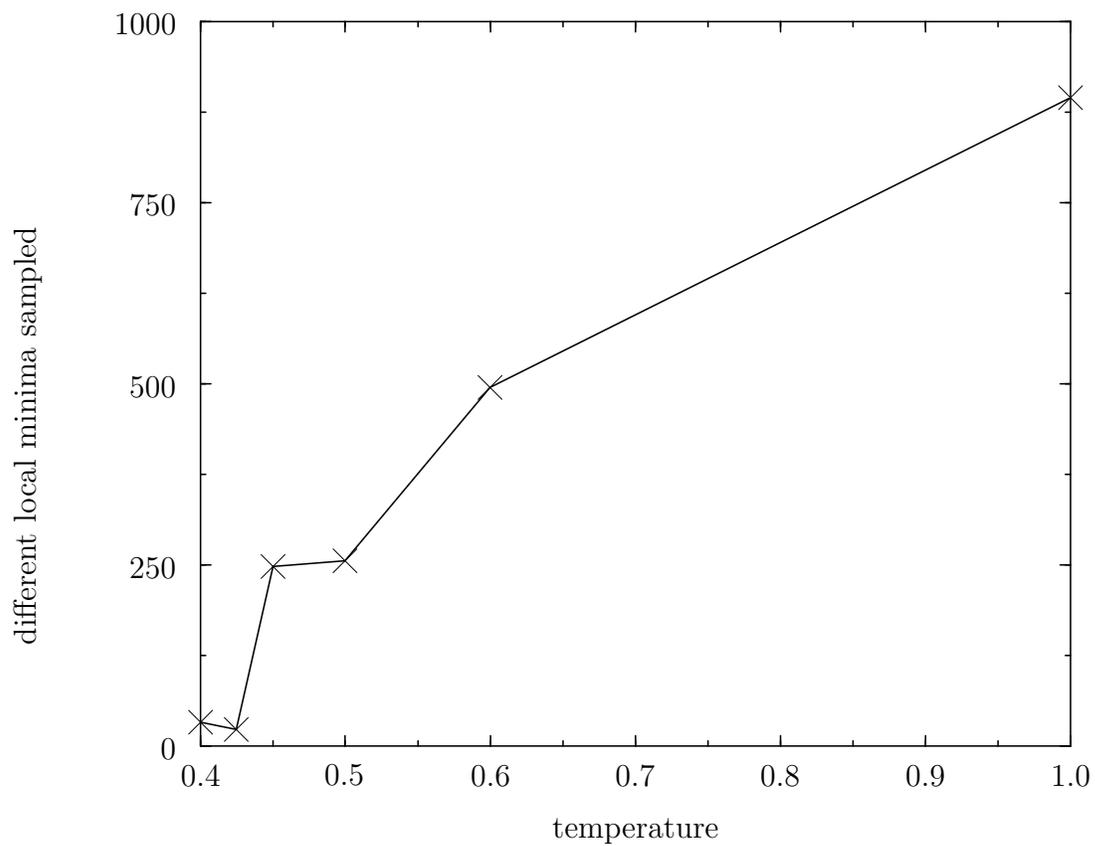}}
\caption{\label{fig:min_v} Number of different local minima visited 
in the last 5000 KMC steps as a function of the temperature.}
\end{figure}

A qualitative change in behaviour at low temperature is clearly 
seen in Figure \ref{fig:min_v}, 
where we plot the number of different local minima (including permutational isomers) visited 
in the last 5000 KMC steps of each run.
The number of different local minima visited declines steadily with temperature until
around $T=0.45$, when it drops by more than an order of magnitude.

\begin{figure}[htbp]
\psfrag{0}[cr][cr]{$0.0$}
\psfrag{0.5}[cr][cr]{$0.5$}
\psfrag{1}[cr][cr]{$1.0$}
\psfrag{1.5}[cr][cr]{$1.5$}
\psfrag{2}[cr][cr]{$2.0$}
\psfrag{T=1}[Bl][Bl]{$T=1.00$}
\psfrag{T=0.5}[Bl][Bl]{$T=0.50$}
\psfrag{T=0.45}[Bl][Bl]{$T=0.45$}
\psfrag{T=0.425}[Bl][Bl]{$T=0.425$}
\psfrag{T=0.4}[Bl][Bl]{$T=0.400$}
\psfrag{5}[tc][tc]{$5$}
\psfrag{10}[tc][tc]{$10$}
\psfrag{15}[tc][tc]{$15$}
\psfrag{20}[tc][tc]{$20$}
\psfrag{25}[tc][tc]{$25$}
\psfrag{30}[tc][tc]{$30$}
\psfrag{q}[tc][tc]{$q$}
\psfrag{SAAq}[Bc][tc]{$S_{\rm AA}(q)$}
\centerline{\includegraphics[width=14.0cm]{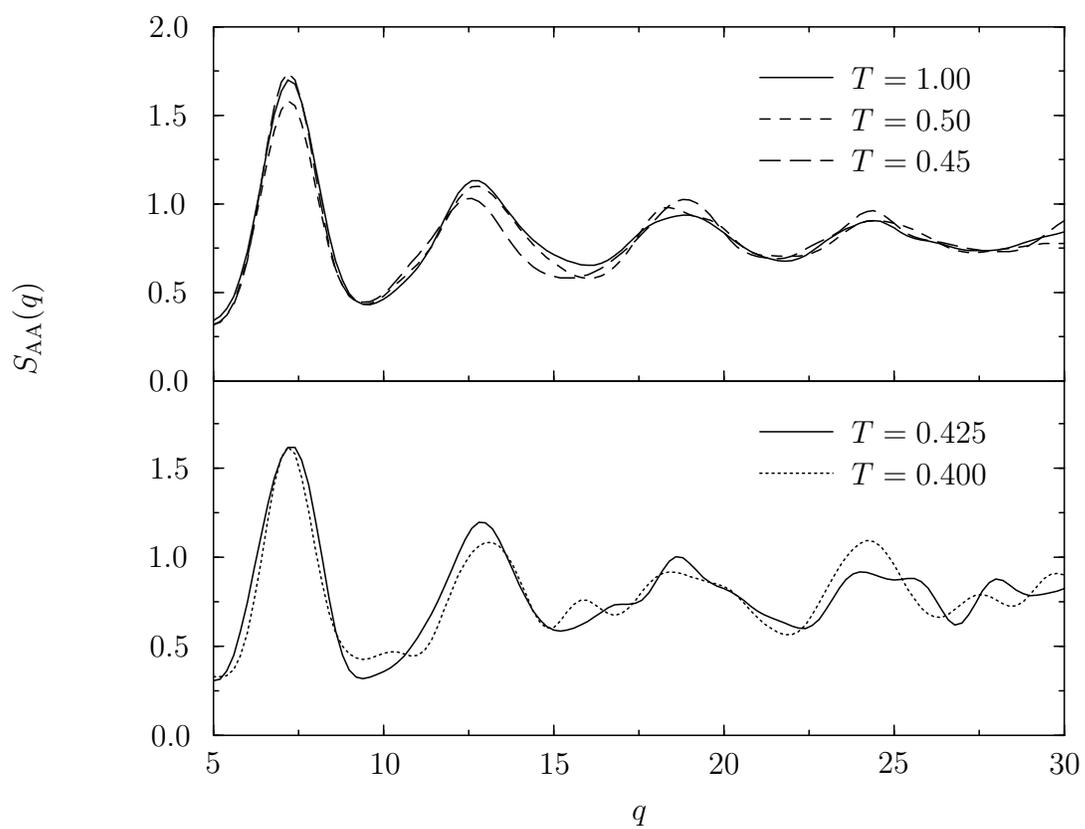}}
\caption{ \label{fig:stru} Partial structure factor for A atoms as a function
of temperature.}
\end{figure}

In order to elucidate whether the system is a supercooled liquid or a glass, we calculated the 
partial structure factor, which encodes information about the typical 
distances between particles and their spatial distribution in a given minimum:
\begin{equation}
S_{\alpha \beta}(q)=\frac 1N \left< \sum\limits_{i=1}^{N_\alpha}\sum\limits_{j=1}^{N_\beta}
e^{i{\bf q}\cdot \left( {\bf r}_i^\alpha-{\bf r}_j^\beta \right) } \right>,
\end{equation}
where the average is over different local minima and fifty different orientations of ${\bf q}$ 
uniformly distributed over the surface of a sphere. 
We have calculated $S_{\alpha \beta}(q)$
for the A atoms over a range of temperatures (Figure \ref{fig:stru}),
averaging over every 25th local minimum 
from the last 5000 minima of each KMC simulation. 
There appears to be little dependence of $S_{\alpha \beta}(q)$ 
on temperature above $T=0.45$, but new peaks appear at lower temperature
and $S_{\alpha \beta}(q)$ attenuates more slowly as a function of $q$.
These results are in good agreement with previous work \cite{koba95b,Kob99}.

\begin{figure}[htbp]
\psfrag{0}[Br][Br]{$0.0$}
\psfrag{0.1}[Br][Br]{$0.1$}
\psfrag{0.2}[Br][Br]{$0.2$}
\psfrag{0.3}[Br][Br]{$0.3$}
\psfrag{Temperature}[tc][tc]{Temperature}
\psfrag{0.4}[tc][tc]{$0.4$}
\psfrag{0.5}[tc][tc]{$0.5$}
\psfrag{0.6}[tc][tc]{$0.6$}
\psfrag{0.7}[tc][tc]{$0.7$}
\psfrag{0.8}[tc][tc]{$0.8$}
\psfrag{0.9}[tc][tc]{$0.9$}
\psfrag{1}[tc][tc]{$1.0$}
\psfrag{<dVmin2>/atom*T2}[Bl][tl]{$(\left<V^2\right>-\left<V\right>^2)/NT^2$}
\centerline{
\includegraphics[width=14.0cm]{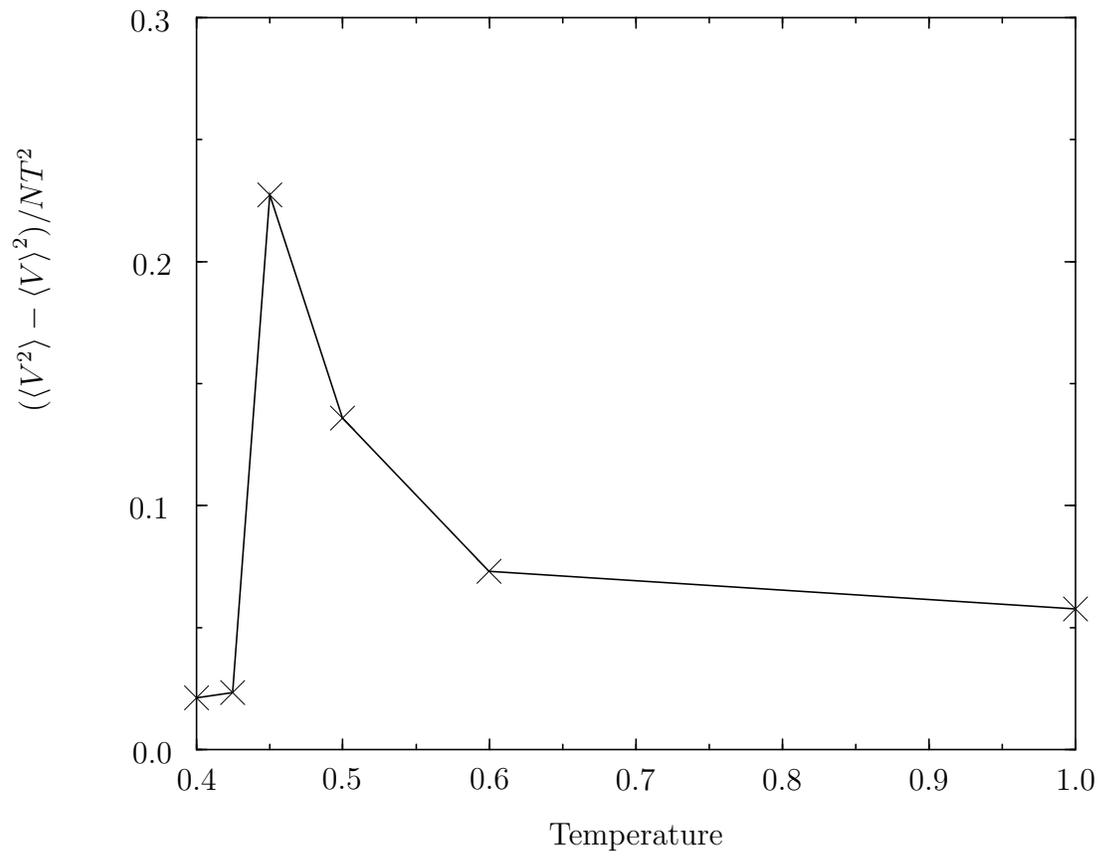}}
\caption{ \label{fig:fluct} Mean square fluctuation of the potential energy divided by
$T^2$ as a function of $T$.}
\end{figure}

\begin{figure}[t]
\psfrag{0}[tc][tc]{$0.0$}
\psfrag{0.1}[tc][tc]{$0.1$}
\psfrag{0.2}[tc][tc]{$0.2$}
\psfrag{0.3}[tc][tc]{$0.3$}
\psfrag{1}[Br][Br]{$1$}
\psfrag{10}[Br][Br]{$10$}
\psfrag{100}[Br][Br]{$100$}
\psfrag{r}[tc][tc]{$r$}
\psfrag{T=1}[Bl][Bl]{$T=1.000$}
\psfrag{T=0.5}[Bl][Bl]{$T=0.500$}
\psfrag{T=0.45}[Bl][Bl]{$T=0.450$}
\psfrag{T=0.425}[Bl][Bl]{$T=0.425$}
\psfrag{T=0.4}[Bl][Bl]{$T=0.400$}
\psfrag{<Number of atoms A moving>}[Bc][Bc]{$N_A(r)$}
\centerline{\includegraphics[width=14.0cm]{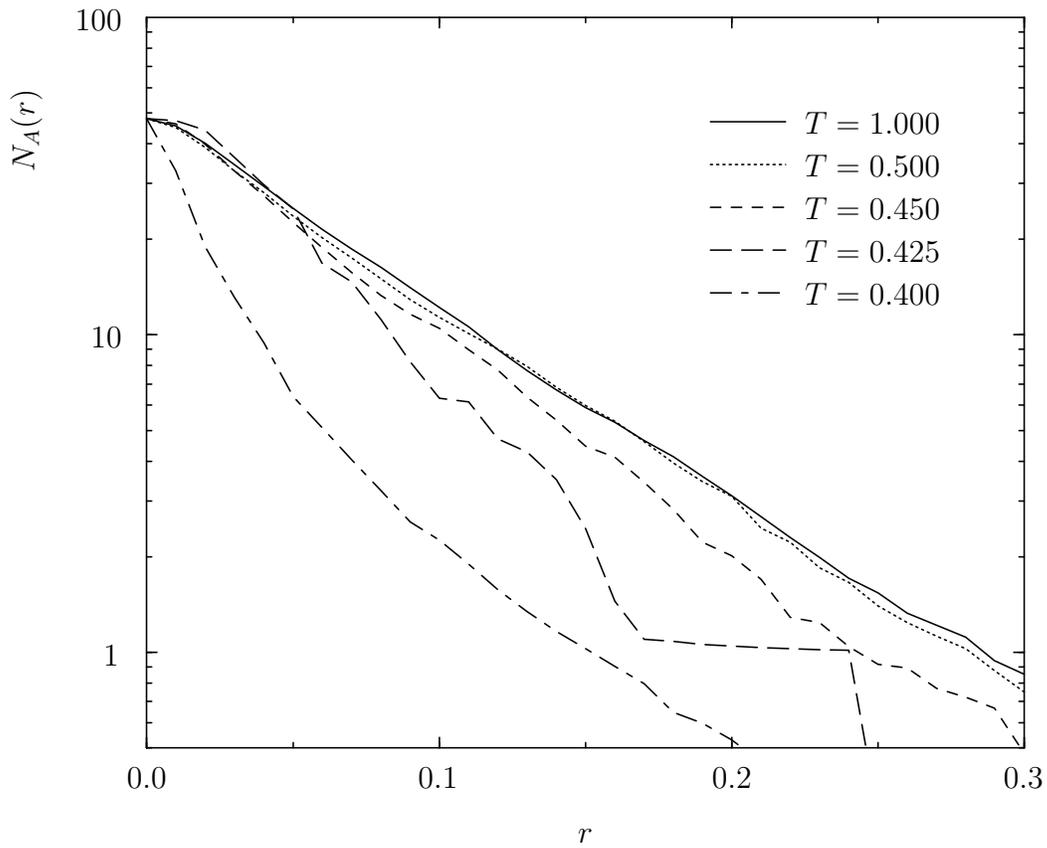}}
\caption{ \label{fig:atom_A} Linear-log plot of the mean number of A atoms moving 
through a distance $r$, $N_A(r)$, for KMC runs at five temperatures.}
\end{figure}

We have also calculated
the variance of the potential energy of the last 5000 local minima
for each run, 
as a first approximation to the configurational component of the heat capacity. 
Figure \ref{fig:fluct} shows that this variance exhibits a peak around $T=0.45$.
Figure \ref{fig:atom_A} shows the mean number of A atoms 
that move more than a distance $r$, again averaged over the final
5000 local minima of each KMC run:
\begin{equation}
N_A(r) = {1\over 5000} \sum_{j=5001}^{10000} \sum_{\alpha=1}^{N_A}
         \Theta\left(|{\bf r}_\alpha(j)-{\bf r}_\alpha(j+1)|-r\right) ,
\end{equation}
where $\Theta$ is the Heaviside step function and ${\bf r}_\alpha(j)$ is the position vector of
atom $\alpha$ in minimum $j$. 
Above $T=0.45$ $N_A(r)$ exhibits an exponential
decay, while at lower temperatures the behaviour is more complicated.
Above $T=0.45$ we can define a characteristic
length in terms of the inverse of the exponential decay rate,
which seems to be only weakly dependent on $T$.

Below $T=0.45$ the absence of a clear exponential decay 
may be due to heterogeneities in the dynamics. The atomic displacements
tend to be shorter and the number of atoms involved is also reduced for both A and B atoms.
These results are in good agreement with Mousseau's ART results for a much larger binary Lennard-Jones system\cite{Mousseau00}
and with the experimental observations of Tang {\it et al.\/}\cite{tanggbjw99} for some
metallic glasses.
Non-ergodic behaviour is again evident from the irregular behaviour of the plots
at low temperature. Only local equilibration is achieved at such temperatures for
KMC runs of the present length, and so the results depend sensitively upon the limited region
that the system then samples. Nevertheless, the sampling should be significantly better
than would be achieved by conventional MD.

\begin{figure}[t]
\psfrag{0}[Br][Br]{$0$}
\psfrag{1}[Br][Br]{$1$}
\psfrag{2}[Br][Br]{$2$}
\psfrag{3}[Br][Br]{$3$}
\psfrag{4}[Br][Br]{$4$}
\psfrag{T=1}[Bl][Bl]{$T=1.0$}
\psfrag{T=0.6}[Bl][Bl]{$T=0.6$}
\psfrag{T=0.5}[Bl][Bl]{$T=0.5$}
\psfrag{T=0.45}[Bl][Bl]{$T=0.450$}
\psfrag{T=0.425}[Bl][Bl]{$T=0.425$}
\psfrag{T=0.4}[Bl][Bl]{$T=0.400$}
\psfrag{0.0}[Br][Br]{$0.0$}
\psfrag{0.5}[Br][Br]{$0.5$}
\psfrag{1.0}[Br][Br]{$1.0$}
\psfrag{1.5}[Br][Br]{$1.5$}
\psfrag{Energy}[tc][tc]{Energy}
\centerline{\includegraphics[width=14.0cm]{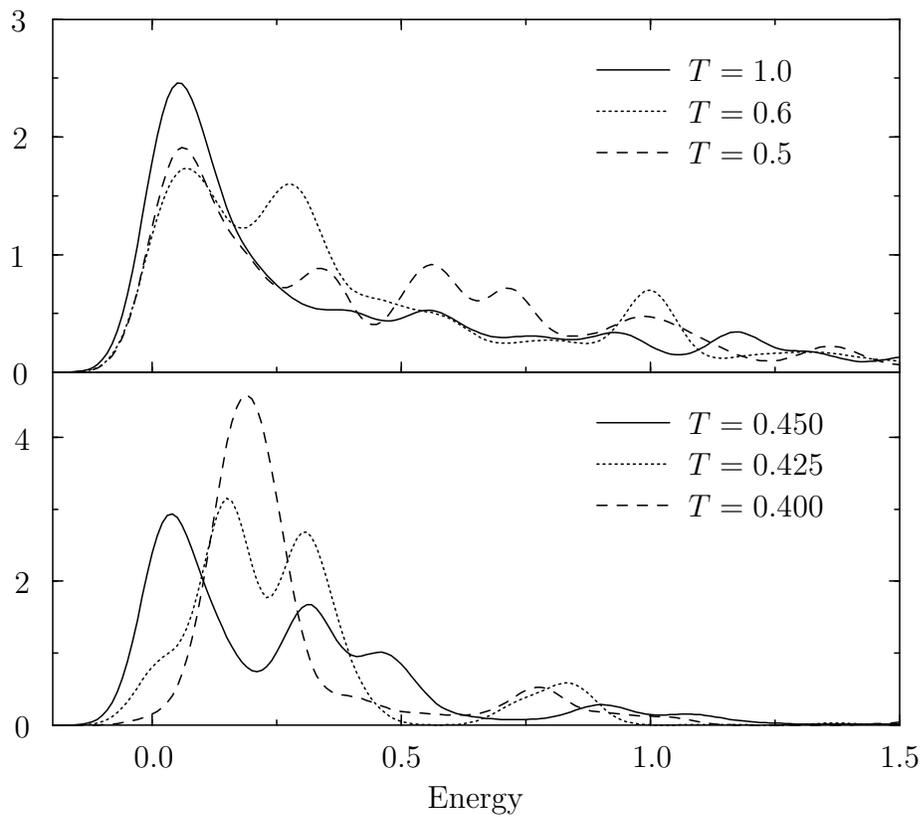}}
\caption{\label{fig:barriers} Normalised barrier distributions for accepted KMC moves as a function of temperature.
Only the barriers corresponding to forward jumps (not backward) are included.}
\end{figure}

Figure \ref{fig:barriers} shows probability distributions for the barriers overcome in the
last 5000 KMC steps as a function of temperature. The distributions were constructed in the
same way as reference \cite{MiddletonW01}, and provide an interesting comparison, since
the canonical KMC sampling is different from the scheme used by Middleton and Wales. The distributions
peak at low energy for the higher temperature runs, in agreement with the previous samples\cite{MiddletonW01}.
However, for the two lowest temperatures there is a distinct shift in the maximum of the distribution
to higher energy. The reason for this discrepancy has not yet been elucidated, but the new results
suggest that the facile rearrangements most often encountered at higher temperature are no longer available
at the lowest temperatures. However, a subsidiary maximum in the distribution is clearly visible around a
barrier height of $0.8$ at $T=0.4$ and $T=0.425$. 
We have previously found that such peaks correspond to diffusive-like processes
in a number of model glasses\cite{MiddletonW01}, where at least one atom changes its nearest-neighbour
coordination shell. For KMC steps where the distance between the local minima is greater than unity,
the forward barrier is always found to exceed 1.5 reduced units.
The main features of the present results are therefore in reasonable agreement
with previous work.

\section{Conclusions}
We have performed numerical simulations of a 
supercooled binary Lennard-Jones liquid using a kinetic Monte Carlo (KMC) approach,
which enables us to perform more extensive sampling at low temperature than the standard
molecular dynamics approach.
Most previous simulation studies have reported some sort of qualitative change
in behaviour as the system approaches the critical temperature of mode-coupling
theory\cite{gotze91,kob97,gotze99}, $T_c\approx0.435$\cite{koba94,koba95a}.
Following Jonsson and Anderson\cite{jonssona88},
Sastry, Debenedetti and Stillinger\cite{sastryds98} found evidence that around
$T=0.45$, above the glass transition,
the height of the barriers separating local minima increases abruptly,
and that the system shows evidence of activated dynamics at about the same temperature.
They referred to the onset of activated dynamics as the `landscape-dominated' regime,
in agreement with the results of Donati, Sciortino and Tartaglia, based on an
instantaneous normal modes picture\cite{donatist00}.
Sastry, Debenedetti and Stillinger also
noted the onset of non-exponential relaxation below about $T=1.0$, and referred to
this region as `landscape-influenced'. Local minima
obtained by quenching from configurations generated at $T=0.5$ were also found to
escape to different local minima much more easily than local minima obtained from
configurations generated at $T=0.4$. The latter result is in good agreement with the
change in the number of different local minima sampled in our calculations,
suggesting that we have indeed sampled sufficient transition states per minimum
to extract meaningful dynamics. This aspect of the KMC calculations will be investigated
more systematically in future work; we intend to check that the results are converged with respect 
to the number of transition state searches and are not affected by any sampling bias
from the search algorithm.

For a BLJ mixture of 50:50 composition Schr\o der {\it et al.}~concluded that the
liquid dynamics could be separated into transitions between minima and vibrational motion,
as originally suggested by Goldstein\cite{goldstein69},
below a temperature around $T_c$\cite{SchroderSDG00}. They also noted correlated
motions of groups of atoms in the same temperature regime, in agreement with results
for an 80:20 BLJ mixture\cite{donatigpkp99}, which suggest that the number of atoms
in the correlated groups grows with decreasing temperature. 
This correlated motion is
not inconsistent with our finding that the number of atoms moving in an elementary step
decreases with temperature, because it involves sequences of such steps.

Most recently, two groups have reported that the typical Hessian index of stationary
points sampled by BLJ systems extrapolates to zero, again around $T_c$\cite{AngelaniDRSS00,BroderixBCZG00}.
Our results show that transition states are still accessible below $T_c$, but support
the general view that the PES sampled by the BLJ system
changes in character somewhere around this point.

In the high temperature regime above about $T=0.45$ the system samples the basins of attraction of 
many different local minima, and the distribution of atomic displacements per
KMC step exhibits exponential decay with a relatively weak temperature dependence.
Hence we can define a characteristic length scale based upon the inverse of the
exponential decay rate. 
In the low temperature regime
the system samples far fewer local minima for the same number of KMC steps,
and the atomic dynamics cannot be described by a single length scale.
We interpret this phenomenon in terms of a multi-funnel 
energy landscape \cite{WalesDMMW00},
where the system is associated with lower lying minima at lower temperature
leading to a higher effective activation energy for diffusion, and non-Arrhenius
dynamics \cite{Bassler87,MillerDW99b,DoyeW99b,WalesD01}.

It is noteworthy that this popular binary Lennard-Jones system does in fact possess a
crystalline phase well separated in energy from the minima sampled by
the supercooled liquid \cite{MiddletonHMW01}. In fact, crystallisation was previously observed by
Kob and Anderson\cite{koba95a} for a system of the same composition but with different
$\epsilon$ parameters.
The present approach can also be used to study dynamical properties such as self-diffusion
or the incoherent intermediate scattering function. This work is now in progress.

\section{Acknowledgements}
J.H-R is grateful to Daniel Alonso for his helpul discussions 
on this manuscript and to the ``Consejer\'{\i}a de Educaci\'on, 
Cultura y Deportes del Gobierno Aut\'onomo de Canarias'' for 
financial support (Proyect No. PI2000/122). 

\def\aciee{Angew.~Chem.~Int.~Ed.~Engl.}
\def\acp{Adv.~Chem.~Phys.}
\def\acr{Acc.~Chem.~Res.}
\def\ac{Acta.~Crystallogr.}
\def\ajp{Am.~J.~Phys.}
\def\am{Adv.~Mater.}
\def\apl{Appl.~Phys.~Lett.}
\def\ap{Ann.~Physik}
\def\Pa{Physica A}
\def\arpc{Ann.~Rev.~Phys.~Chem.}
\def\bbpc{Ber. Bunsenges. Phys. Chem.}
\def\bc{Biochemistry}
\def\cccc{Coll.~Czech.~Chem.~Comm.}
\def\cj{Comput.~J.}
\def\cpc{Comp.~Phys.~Comm.}
\def\cpl{Chem.~Phys.~Lett.}
\def\cp{Chem.~Phys.}
\def\crev{Chem.~Rev.}
\def\dalton{J.~Chem.~Soc., Dalton Trans.}
\def\el{Europhys.~Lett.}
\def\faraday{J.~Chem.~Soc., Faraday Trans.}
\def\fartrans{J.~Chem.~Soc., Faraday Trans.}
\def\fdisc{J.~Chem.~Soc., Faraday Discuss.}
\def\ic{Inorg.~Chem.}
\def\ijmpc{Int.~J.~Mod.~Phys.~C}
\def\ijqc{Int.~J.~Quant.~Chem.}
\def\jacers{J. Am. Ceram. Soc.}
\def\jacs{J.~Am.~Chem.~Soc.}
\def\jap{J.~Appl.~Phys.}
\def\jas{J.~Atmos.~Sci.}
\def\jcc{J.~Comp.~Chem.}
\def\jce{J.~Chem.~Ed.}
\def\jcis{J.~Colloid Interface Sci.}
\def\jcp{J.~Chem.~Phys.}
\def\jcscc{J.~Chem.~Soc., Chem.~Commun.}
\def\jcsft{J.~Chem.~Soc., Faraday Trans.}
\def\jetp{J.~Exp.~Theor.~Phys.~(Russia)}
\def\jmc{J.~Math.~Chem.}
\def\jmsp{J.~Mol.~Spec.}
\def\jmst{J.~Mol.~Structure}
\def\jncs{J.~Non-Cryst.~Solids}
\def\jpa{J.~Phys.~A}
\def\jpca{J.~Phys.~Chem.~A}
\def\jpcb{J.~Phys.~Chem.~B}
\def\jpcm{J.~Phys.~Condensed Matter.}
\def\jpcssp{J.~Phys.~C: Solid State Phys.}
\def\jpcs{J.~Phys.~Chem.~Solids.}
\def\jpc{J.~Phys.~Chem.}
\def\jpfmp{J.~Phys.~F, Metal Phys.}
\def\jpsj{J.~Phys.~Soc.~Jpn.}
\def\jsp{J.~Stat.~Phys.}
\def\mg{Math.~Gazette}
\def\molphys{Mol.~Phys.}
\def\molp{Mol. Phys.}
\def\mrsb{Mater.~Res.~Soc.~Bull.}
\def\msr{Mater.~Sci.~Rep.}
\def\nat{Nature}
\def\njc{New J.~Chem.}
\def\pac{Pure.~Appl.~Chem.}
\def\phys{Physics}
\def\pla{Phys.~Lett.~A}
\def\phm{Philos. Mag.}
\def\pma{Philos.~Mag.~A}
\def\pmb{Philos.~Mag.~B}
\def\pml{Philos.~Mag.~Lett.}
\def\pnasu{Proc.~Natl.~Acad.~Sci.~USA}
\def\pnas{Proc.\ Natl.\ Acad.\ Sci.\  USA}
\def\pra{Phys.~Rev.~A}
\def\prbcm{Phys.~Rev.~B}
\def\prb{Phys.~Rev.~B}
\def\prc{Phys.~Rev.~C}
\def\prd{Phys.~Rev.~D}
\def\prep{Phys.~Reports}
\def\pre{Phys.~Rev.~E}
\def\prl{Phys.~Rev.~Lett.}
\def\prsa{Proc.~R.~Soc.~A}
\def\pr{Phys.~Rev.}
\def\psfg{Proteins: Struct., Func.~and Gen.}
\def\pssb{Phys.~State Solidi B}
\def\pss{Phys.~State Solidi}
\def\rmp{Rev.~Mod.~Phys.}
\def\rpp{Rep.~Prog.~Phys.}
\def\sci{Science}
\def\ss{Surf.~Sci.}
\def\tca{Theor.~Chim.~Acta}
\def\tetra{Tetrahedron}
\def\zfpd{Z.~Phys.~D}
\def\zpb{Z.~Phys.~B.}
\def\zpc{Z.~Phys.~Chem.}
\def\zpdamc{Z.~Phys.~D}
\def\zpd{Z.~Phys.~D}

\newpage

\end{document}